\documentclass[aps,prl,twocolumn,showpacs,superscriptaddress,footinbib]{revtex4-1}
\usepackage{graphicx}% Include figure files
\usepackage{xcolor}
\usepackage{dcolumn}% Align table columns on decimal point
\usepackage{bm}% bold math
\usepackage{enumerate}
\usepackage{amsmath}

\usepackage[normalem]{ulem}  % \sout{old text} for strikeout
\begin{document}

%\preprint{APS/123-QED}

\title{The boson number hypothesis and the boson number odd-even effect in $^{196-204}$Hg}

\author{Tao Wang}
\email{suiyueqiaoqiao@163.com}
\affiliation{College of Physics, Tonghua Normal University, Tonghua 134000, People's Republic of China}

\author{Chun-xiao Zhou}
\email{zhouchunxiao567@163.com}
\affiliation{College of Mathematics and Physics Science, Hunan University of Arts and Science, Changde 415000, People's Republic of China}

\author{Lorenzo Fortunato}
\email{fortunat@pd.infn.it}
\affiliation{Dipartimento di Fisica e Astronomia  ``G.Galilei''-Universit\`{a} di Padova, via Marzolo 8, I-35131 Padova, Italy }
\affiliation{INFN-Sez.di Padova, via Marzolo 8, I-35131 Padova, Italy}

\date{\today}

\begin{abstract}
In the SU3-IBM the oblate shape is described by the \textrm{SU(3)} third-order Casimir operator in the large-$N$ limit. However for finite $N$, this interaction can produce a boson number odd-even effect. In this Letter, we find that, the unique odd-even effect really exists in the nuclei $^{196-204}$Hg. This finding implies that realistic low-lying excitations are sensitive to certain boson number $N$. The boson number hypothesis is verified for the first time since the advent of the interacting boson model. This also proves the accuracy and validity of the SU3-IBM directly. The SU(3) symmetry and the higher-order interactions are both indispensable for understanding the nuclear quadrupole deformations.
\end{abstract}

\maketitle

The interacting boson model (IBM), proposed by Arima and Iachello 50 years ago \cite{Iachello75, Iachello87}, is an influential algebraic model of the atomic nuclei. The main hypothesis at its foundation  is that the low-lying collective nuclear states are the totally symmetric representation $[N]$ of the SU(6) group that arises when the nucleus is built in terms of $N$ valence bosons with either $s$  ($L=0$) or $d$ ($L=2$) character. Recently, an extension of the interacting boson model with \textrm{SU(3)} higher-order interactions (SU3-IBM) was proposed by one of the authors (T. Wang) \cite{Wang20,Wang22}, which emphasizes the SU(3) symmetry \cite{Kota20} and the higher-order interactions \cite{Isacker81,Isacker85}. The large-$N$ limit of this new model was first discussed by one of the authors (L. Fortunato) and his collaborators \cite{Fortunato11}.
In the SU3-IBM, the SU(3) symmetry is suggested to dominate over the onset of quadrupole deformations, and only the U(5) limit and the SU(3) limit are included. So, in a sense, the SU3-IBM is based on a very restrictive set of hypotheses. In the SU(3) limit, the second-order Casimir operator $\hat{C}_{2}[\textrm{SU(3)}]$ can describe prolate shapes, the third-order Casimir operator $\hat{C}_{3}[SU(3)]$ can describe oblate shapes, and a combination of these two with the square of the second-order Casimir operator $\hat{C}_{2}^{2}[\textrm{SU(3)}]$ can produce triaxial shapes. When the $d$ boson number operator in the U(5) limit is considered, the emerging $\gamma$-softness can appear \cite{Wang22,WangPt}, without invoking the O(6) limit \cite{Iachello78,Casten78}.

\begin{figure}[tbh]
\includegraphics[scale=0.29]{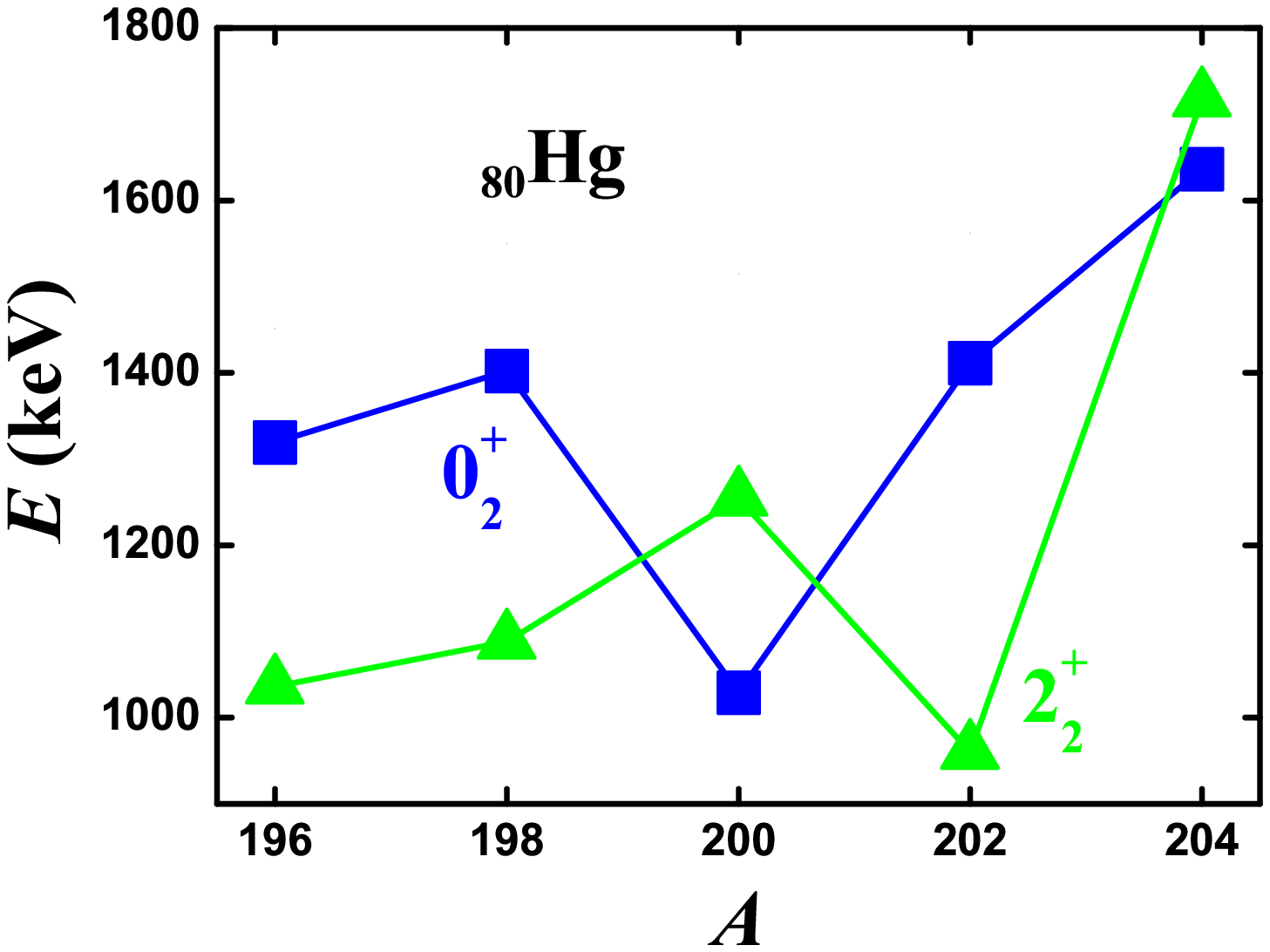}
\caption{Boson number odd-even effect in $^{196-204}$Hg \cite{ensdf}.}
\end{figure}

So far, this new model has been used to explain the B(E2) anomaly \cite{Wang20,Zhang22,wangtao,Zhang24,Pan24,Teng25,Zhang25,Wang25}, to describe the normal states in $^{110}$Cd \cite{Wang22,Wang24}, to describe the prolate-oblate asymmetric shape phase transition in Hf-Hg region \cite{wang23}, to describe the E(5)-like spectra in $^{82}$Kr \cite{Zhou23}, and to better describe the $\gamma$-soft spectra in $^{196}$Pt \cite{WangPt,ZhouPt}. Recently, two new important findings are: (1) $^{106}$Pd is a typical spherical-like nucleus \cite{WangPd}, whose spectra was proposed to resolve the Cd puzzle \cite{Heyde11,Heyde16,Garrett18}, and (2) it can successfully describe the rigid triaxiality of the large deformed nucleus $^{166}$Er \cite{ZhouEr}, which was proposed by Otsuka \emph{et al.} \cite{Otsuka19,Otsuka21,Otsuka24}. All these new results imply that the SU3-IBM can indeed provide an accurate Hamiltonian to describe the properties of low-lying collective excitations.

The key difference between previous IBM and the SU3-IBM is the description of the oblate shape. In previous IBM, the oblate shape is described by the $\overline{\textrm{SU(3)}}$ symmetry in the large-$N$ limit \cite{Smirnov98}. For finite $N$, its spectra is a mirror image of the ones in the SU(3) symmetry for the same boson number $N$ \cite{Jolie03,Wang08} and the representation of its ground state is $(0,2N)$. Thus for different $N$, the energy spectra are similar. In the SU3-IBM, the oblate shape is induced  by the SU(3) third-order Casimir operator in the large-$N$ limit. For finite $N$, the representation of the ground state is $(0,N)$ for the even boson number while $(2,N-1)$ for the odd boson number \cite{Zhang12}. The quantum numbers of the angular momentum of the bandheads of the lowest three energy bands are (0,0,2) for even boson number or (0,2,0) for odd boson number (see Fig. 2 in \cite{Zhang12}). Thus the SU3-IBM predicts a unique boson number odd-even effect for the $0_{2}^{+}$ and $2_{2}^{+}$ states when the boson number changes gradually. It should be noticed that, in the large-$N$ limit, the two representation $(0,N)$ and $(2, N-1)$ both corresponds to the oblate shape. However for finite $N$, the spectra for odd or even boson number are significantly different. If this odd-even effect can be found in realistic nuclei, the validity of the SU3-IBM will be directly confirmed.

It should be noticed that if the boson number odd-even effect is verified, the fundamental hypothesis, that the boson number $N$ is half the number of valence nucleons, can also be directly established. It should be stressed that this odd-even effect can not be found in the large-$N$ limit of the SU3-IBM (see Fig. 12 in \cite{Fortunato11}), previous IBM \cite{Wang08,Werner01,Jolie02,Warner02,Jolie031,Bonatsos24} and other theories on shape evolution \cite{Nomura13,Heyde14,Li21,Utsuno24}.

 $^{196-204}$Hg provide an ideal laboratory to study the boson number odd-even effect. If the SU3-IBM is correct, this odd-even effect should be found in these isotopes. When looking for evidence, we find that, as an anomalous phenomenon, this effect has been observed by Bernards \emph{et al.} \cite{Bernards131,Bernards132}. Fig. 1 shows the evolutional behaviors of the $0_{2}^{+}$ and $2_{2}^{+}$ states in $^{196-204}$Hg. Bernards \emph{et al.} found that the energy of the $0_{2}^{+}$ state in $^{200}$Hg is much lower than the neighboring ones. Obviously, this is just the expected phenomenon that the SU3-IBM predicts. For completely understand these evolutional behaviors, other supplementary interactions should be also considered except for the SU(3) third-order Casimir operator.

Now we explain this effect with the SU3-IBM. The Hamiltonian is as follows \cite{Zhang22,Wang24}
\begin{eqnarray}
\hat{H}&=&c\{(1-\eta)\hat{n}_{d}+\eta[-\frac{\hat{C}_{2}[SU(3)]}{2N}+\alpha\frac{\hat{C}_{3}[SU(3)]}{2N^{2}}   \nonumber\\
&&+\beta\frac{\hat{C}_{2}^{2}[SU(3)]}{2N^{3}}+\gamma\frac{\Omega}{2N^{2}}+\delta\frac{\Lambda}{2N^{3}}]\},
\end{eqnarray}
where $\eta$, $\alpha$, $\beta$, $\gamma$ and $\delta$ are five fitting parameters to determine the structure of the low-lying spectra and $0\leq \eta \leq 1$. Generally, $\alpha\geq 0$ and $\beta \geq 0$. $c$ is a global energy scale parameter. $\hat{n}_{d}$ is the $d$ boson number operator. $\Omega$ is $[\hat{L}\times \hat{Q} \times \hat{L}]^{(0)}$ and $\Lambda$ is $[(\hat{L}\times \hat{Q})^{(1)} \times (\hat{L} \times \hat{Q})^{(1)}]^{(0)}$. $\hat{Q}$ is the SU(3) quadrupole operator. These two quantities can be derived from the SU(3) mapping of the rigid triaxial rotor \cite{Isacker00,zhang14}. The $-\hat{C}_{2}[SU(3)]$, $\hat{C}_{3}[SU(3)]$ and $\hat{C}_{2}^{2}[SU(3)]$ can generate any quadrupole deformation for the ground state with the SU(3) irreducible representation $(\lambda,\mu)$ in the SU(3) limit \cite{Isacker00,zhang14}.

\begin{figure}[tbh]
\includegraphics[scale=0.29]{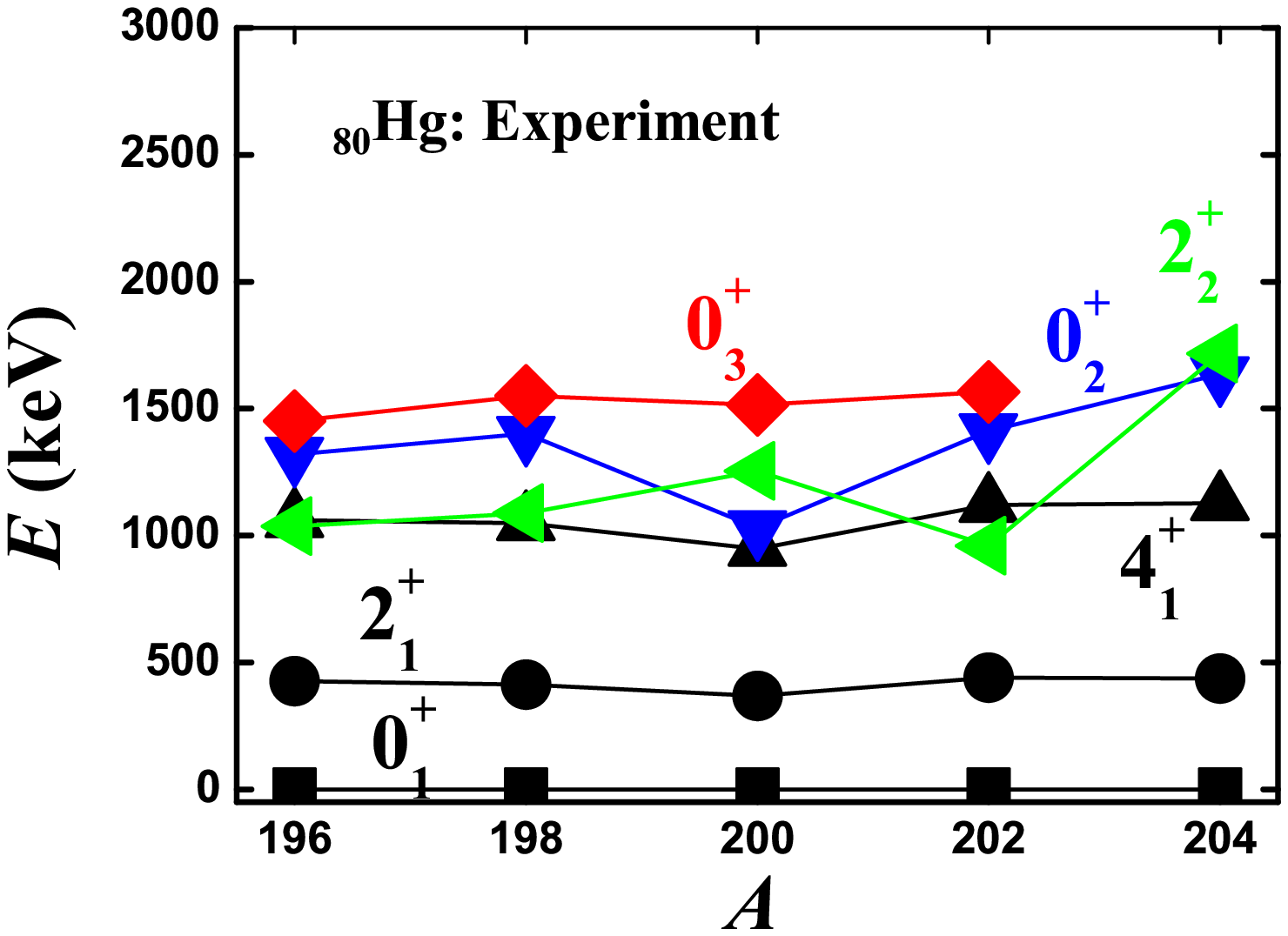}
\includegraphics[scale=0.29]{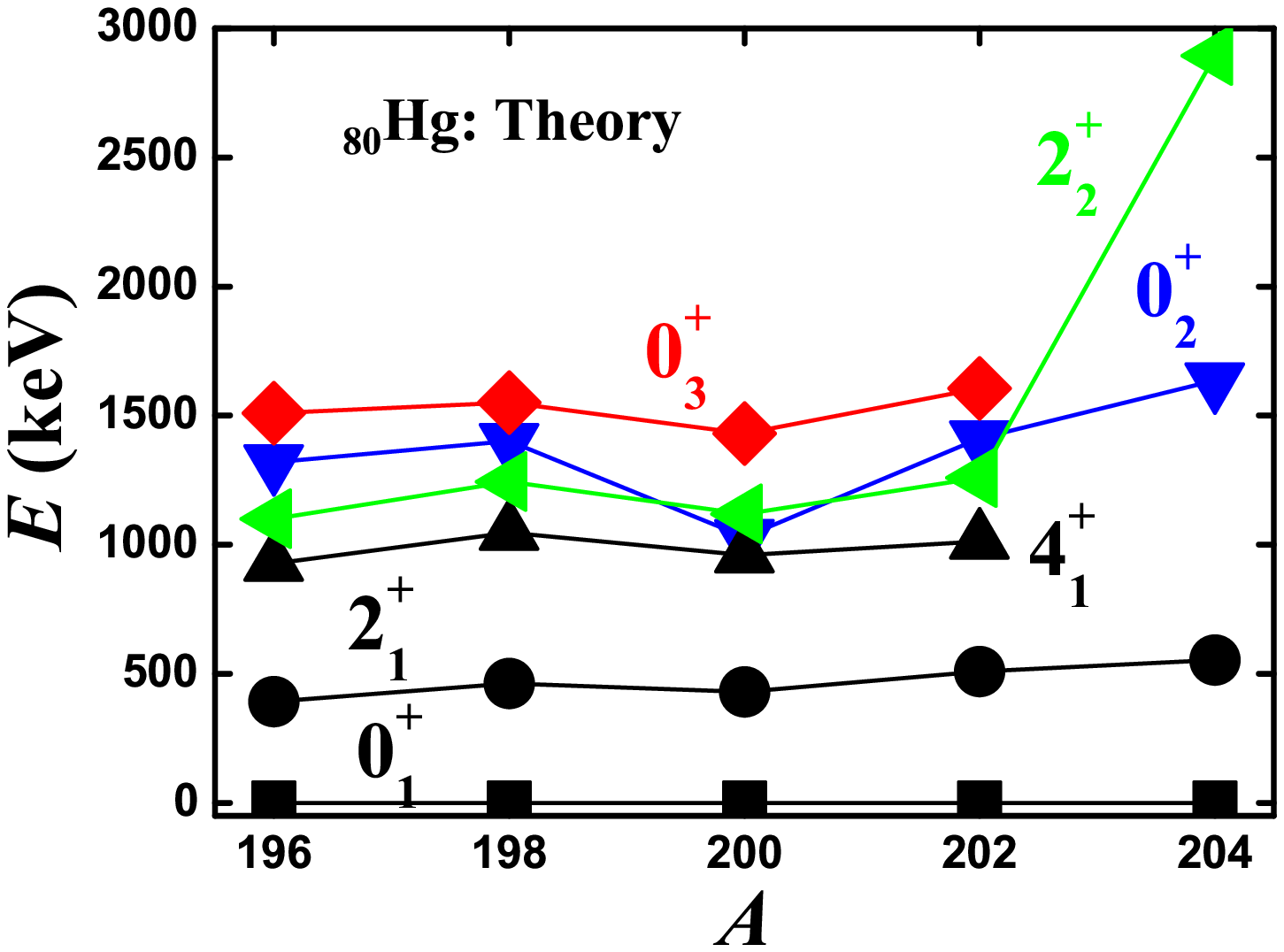}
\caption{Experimental data (top) \cite{ensdf} and the theoretical results (bottom) of the $0_{1}^{+}$, $2_{1}^{+}$, $4_{1}^{+}$, $0_{2}^{+}$, $2_{2}^{+}$ and $0_{3}^{+}$ states in $^{196-204}$Hg.}
\end{figure}

In Ref. \cite{wang23}, the prolate-oblate asymmetric shape phase transition was found in Hf-Hg region, and the degree of the deformation of the prolate shape is nearly twice that of the oblate shape. The qualitative evolution trends can be reproduced well by the Hamiltonian (1) even when $\beta,\gamma,\delta=0$. This implies that the SU3-IBM is useful for the description of the oblate shape. Thus a quantitative fit is needed, and other three SU(3) higher-order interactions should be considered.

\begin{figure}[tbh]
\includegraphics[scale=0.29]{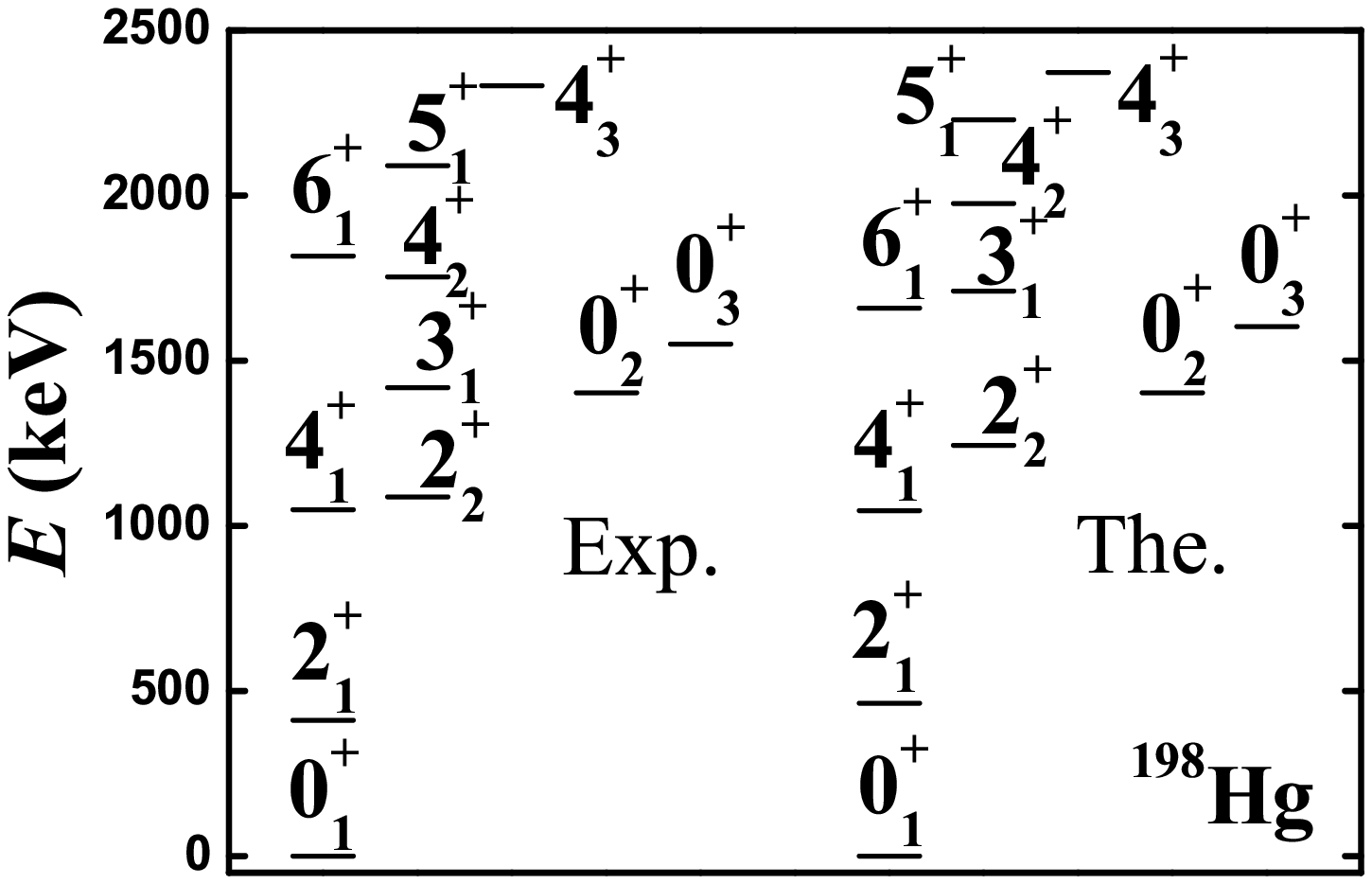}
\includegraphics[scale=0.29]{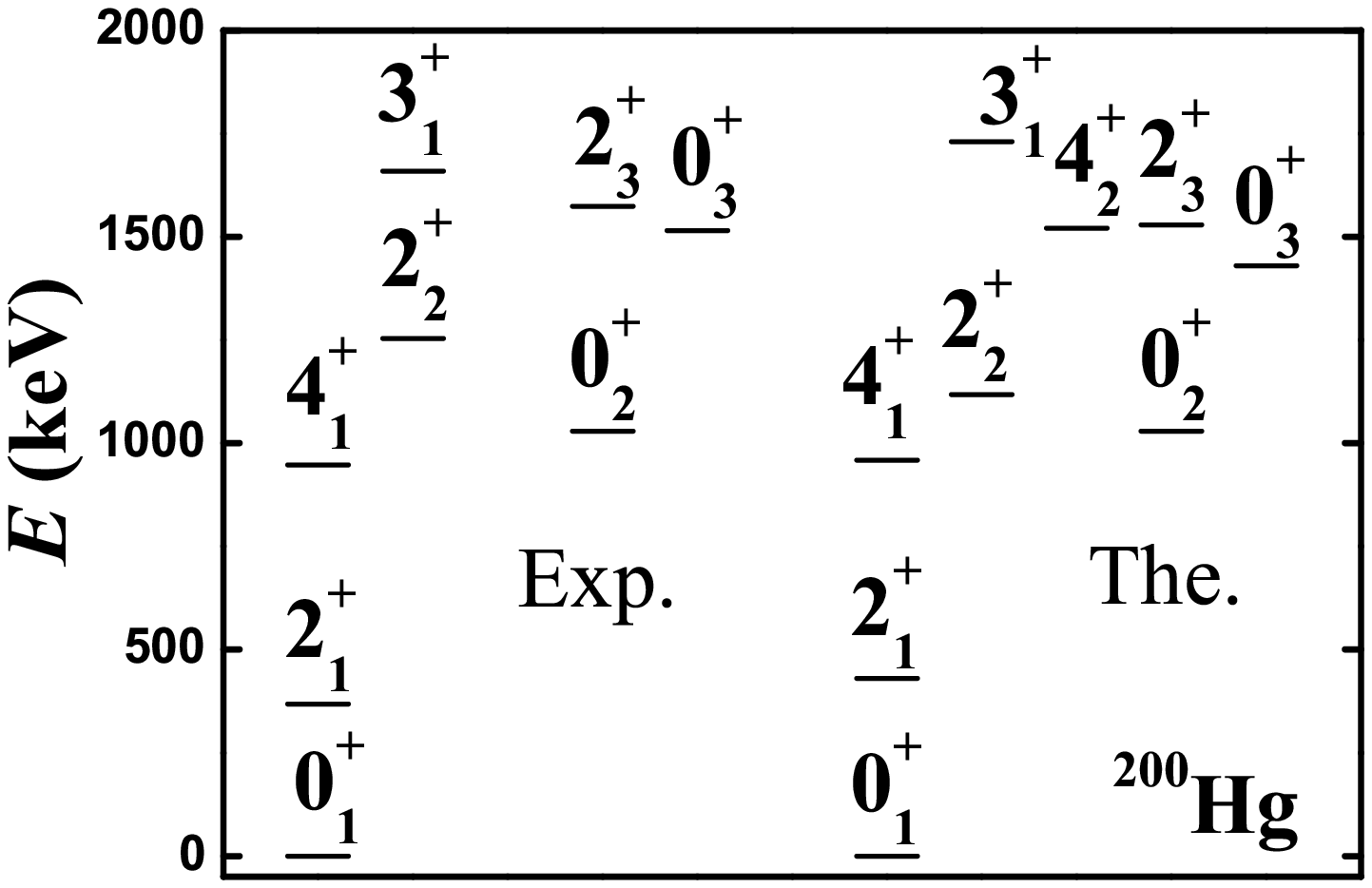}
\includegraphics[scale=0.29]{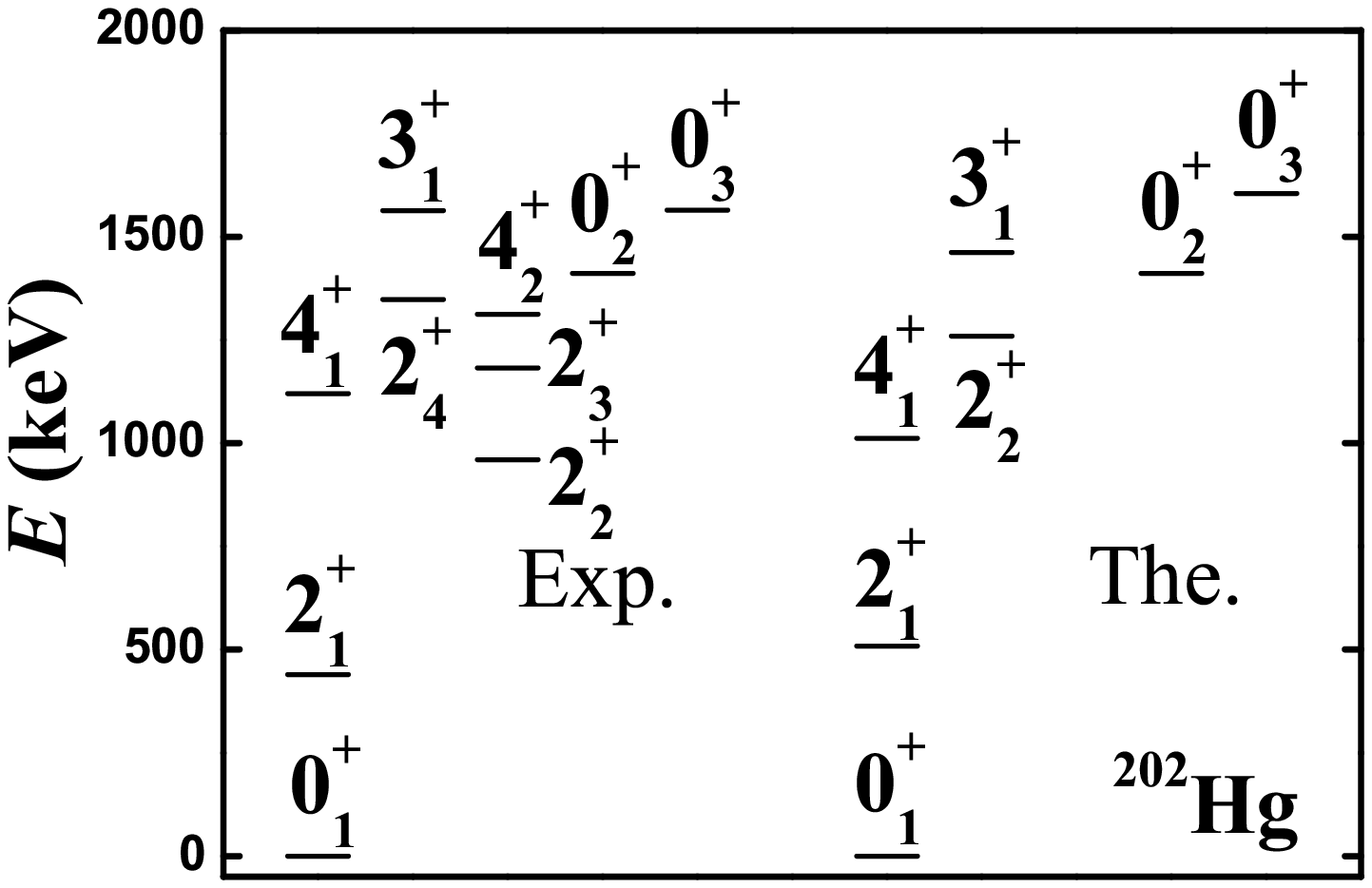}
\caption{Experimental data  \cite{ensdf} and the theoretical results of the low-lying states in $^{198-202}$Hg. Experimental data of $^{198}$Hg is also obtained from Ref. \cite{Chakraborty19}.}
\end{figure}

From previous analysis, it can be seen that, in order to understand the evolutional behaviors in Fig. 1, the main interaction in Hamiltonian (1) is the SU(3) third-order Casimir operator. The boson number for $^{196-204}$Hg are 6,5,4,3,2. The SU(3) representations of the ground state of this third-order operator for $N=2,3,4,5,6$ are (0,2), (2,2), (0,4), (2,4), (0,6), respectively. Thus for only this third-order interaction, the quantum number of the angular momentums of the bandheads of the first excited band are 0,2,0,2,0 for $N=2,3,4,5,6$, which predicts the odd-even effect. Compared with Fig. 1, it is clear that the quantum number of the angular momentums are correct except for the one in $^{196}$Hg, which is 2. This implies that other SU(3) higher-order interactions are also important for understanding the realistic evolutional behaviors, which can change the quantum number in $^{196}$Hg from 0 to 2 while keep the same in heavier nuclei. Thus, in this Letter, we show that the main interaction is the SU(3) third-order Casimir operator for the explanation of the boson number odd-even effect.

\begin{table}[tbh]
\caption{\label{table:expee}  Experimental \cite{ensdf} and theoretical results for the absolute $B(E2)$ values in W.u. for $E2$ transitions from the low-lying states in $^{196-204}$Hg with effective charge $e=2.395,2.542,2.829,2.801,3.471$ (W.u.)$^{1/2}$, respectively. Experimental data in central parentheses are obtained from Ref. \cite{Obaizola19}.}
\setlength{\tabcolsep}{4.0mm}{
\begin{tabular}{cccccccc}
\hline
\hline
Nucleus~~~~$L_{i}  ~~L_{f}    $ &Experiment                      \ &Theory           \\
 \hline
$^{196}$Hg~~~~~~~~$2_1^+  ~~ 0_1^+   $ & 33.3(12)[36(7)]                \ & 33.3               \\
~~~~~~~~~~~~~~~~$4_1^+  ~~ 2_1^+   $ & [20(15)]              \ & 42.7                \\
 \hline
$^{198}$Hg~~~~~~~~$2_1^+   ~~  0_1^+  $ & 28.8(4)[28(4)]     \ & 28.8                      \\
~~~~~~~~~~~~~~~~$4_1^+  ~~  2_1^+  $ & 10.8(5);43(2)[$>$16]                  \ & 33.8              \\
~~~~~~~~~~~~~~~~$2_2^+  ~~ 2_1^+   $ & 0.63(8)                  \ & 32.7               \\
~~~~~~~~~~~~~~~~$2_2^+   ~~ 0_1^+   $ & 0.0216(4)                   \ & 3.44               \\
~~~~~~~~~~~~~~~~$6_1^+  ~~ 4_1^+   $ & 9.0(8)               \ & 22.9               \\
 \hline
$^{200}$Hg~~~~~~~~$2_1^+  ~~ 0_1^+   $ & 24.56(22)[26(2)]                    \ & 24.56                \\
~~~~~~~~~~~~~~~~$4_1^+   ~~ 2_1^+   $ & 38.2(17)[20(9)]            \ & 22.9                  \\
~~~~~~~~~~~~~~~~$0_2^+   ~~  2_1^+  $ & 8(4)               \ & 0.65                   \\
~~~~~~~~~~~~~~~~$2_2^+  ~~ 0_2^+   $ & 10.2(24)[4(3)]        \ & 26.0              \\
~~~~~~~~~~~~~~~~$2_2^+  ~~ 4_1^+   $ & 1.4(4)[0.6(5)]                 \ & 0.06               \\
~~~~~~~~~~~~~~~~$2_2^+   ~~  2_1^+  $ & 2.2(5)[1.0(8)]               \ & 19.1                \\
~~~~~~~~~~~~~~~~$2_2^+    ~~  0_1^+  $ & 0.23(5)[0.10(8)]                     \ & 3.85                \\
 \hline
$^{202}$Hg~~~~~~~~$2_1^+  ~~  0_1^+  $ & 17.34(14)                  \ & 17.34               \\
~~~~~~~~~~~~~~~~$2_2^+   ~~  2_1^+  $ & 5.6(15)                  \ & 23.6            \\
~~~~~~~~~~~~~~~~$2_2^+  ~~  0_1^+  $ & 0.087(21)                    \ & 1.42               \\
~~~~~~~~~~~~~~~~$4_1^+  ~~  2_1^+  $ & 26.5(8)                 \ & 14.8             \\
 \hline
$^{204}$Hg~~~~~~~~$2_1^+   ~~  0_1^+  $ & 11.96(9)                   \ & 11.96               \\
\hline
\hline
\end{tabular}}
\end{table}

To clarify that this odd-even effect is \emph{mainly} due to the change of the boson number and sensitive to certain boson number $N$, here we fix these parameters $\eta=0.4$, $\alpha=1.65$, $\beta=0.05$, $\gamma=-1.9$ and $\delta=-12.0$, which determine the structure of the spectra. To make the energy of the $0_{2}^{+}$ state be equal to the experimental data, the parameter $c$ are 733.5 keV, 743.1 keV, 602.0 keV, 548.6 keV, 613.2 keV for $^{196-204}$Hg, respectively. Thus for this specific Hamiltonian, the only variable is the boson number $N$. In previous studies with the IBM, this kind of fitting did not exist except for a recent paper using the SU3-IBM for the normal states in $^{108-116}$Cd \cite{Wang24}, which reproduces the anomalous trends of the quadrupole moment of the $2_{1}^{+}$ states.

Fig. 2 shows the experimental data (top) and the fitting results (bottom) of the $0_{1}^{+}$, $2_{1}^{+}$, $4_{1}^{+}$, $0_{2}^{+}$, $2_{2}^{+}$ and $0_{3}^{+}$ states in $^{196-204}$Hg. Clearly the theoretical calculations reproduce the evolutional behaviors of the experimental results at a high level except that the theoretical values of the $2_{2}^{+}$ states are somewhat quantitatively worse. The theoretical value of the $2_{2}^{+}$ state in $^{204}$Hg is much larger than the one of the $0_{2}^{+}$ state. Obviously, the theoretical $2_{2}^{+}$ and $0_{2}^{+}$ states belong to the same band while the experimental ones are not so. Experimental odd-even effect can be really explained by the SU3-IBM with a single Hamiltonian. Interestingly, the quantum number of the angular momentum of the bandhead of the first excited band in $^{196}$Hg is 2 while other ones in $^{198-204}$Hg are not changed. Obviously, these fitting results are sensitive to the specific boson number $N$. Adding one boson $N+1$ or reducing one boson $N-1$ cannot give the same result \cite{Wang24}.

\begin{figure}[tbh]
\includegraphics[scale=0.29]{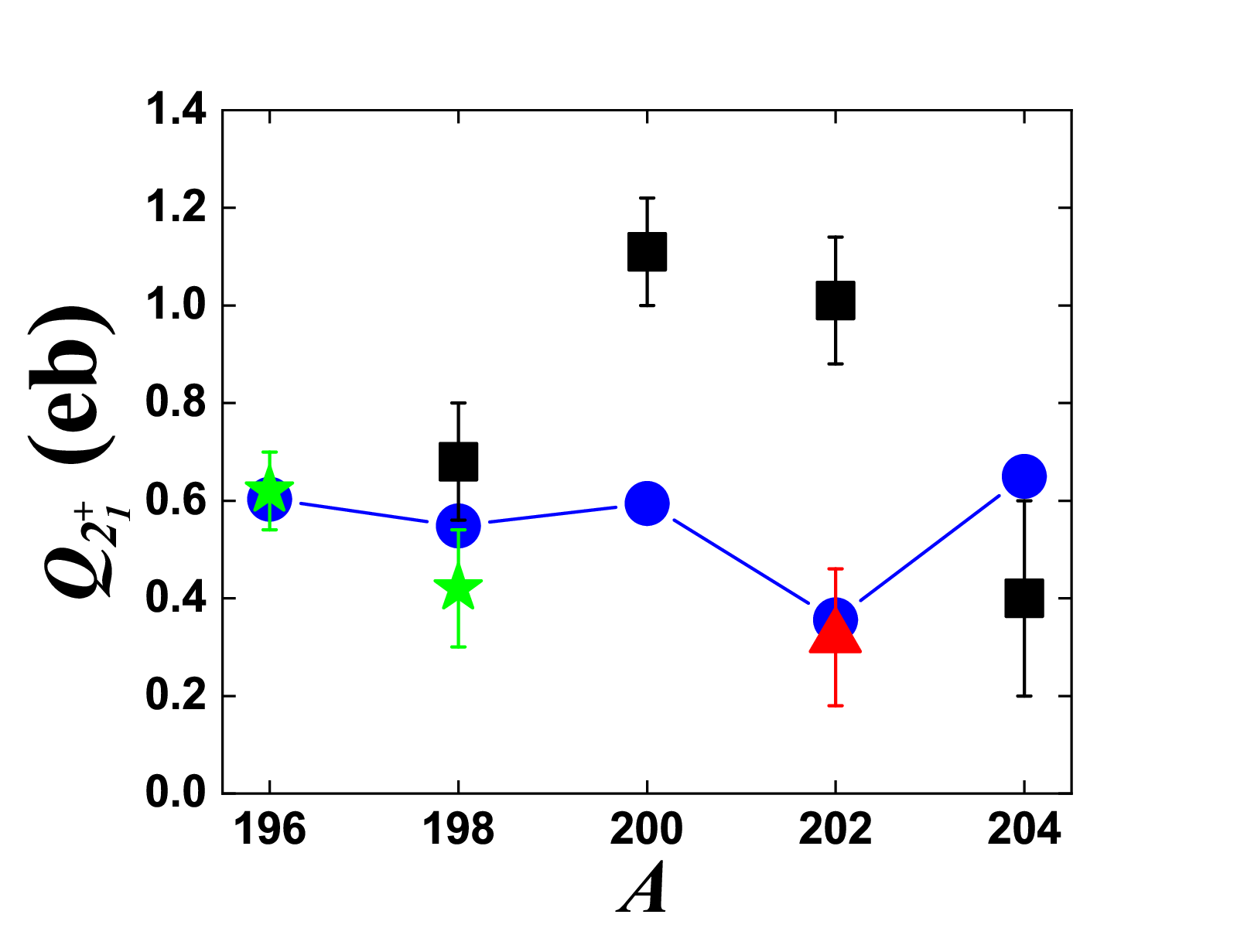}
\caption{Experimental (Black square) \cite{ensdf} and theoretical (blue sphericity) $Q_{2_{1}^{+}}$ values in $^{196-204}$Hg. Red triangle presents another experimental result for $^{202}$Hg \cite{ensdf}. Green stars present the experimental values for $^{196,198}$Pt \cite{ensdf}.}
\end{figure}

To better illustrate the accuracy of the SU3-IBM, Fig. 3 shows the comparisons between experimental data and theoretical results in $^{198-202}$Hg. For the three nuclei, the theory reproduce the low-lying states at a nearly perfect level. For $^{198}$Hg, the experimental and theoretical values of the $4_{3}^{+}$ state are nearly the same. For $^{200}$Hg, at the level of the $2_{3}^{+}$ state, the theory shows that there exists a $4_{2}^{+}$ state. We expect that the future experiment can find it. For $^{202}$Hg, the boson number is 3, less experimental data can be explained. Experimentally a lower $4_{2}^{+}$ state exists. Thus some extensions of the SU3-IBM are needed, such as distinguishing between protons and neutrons \cite{Iachello87}, including the $g$ bosons with $L=4$ \cite{Iachello87}, and including the single particle excitation.

Table I presents the experimental and theoretical B(E2) values among the low-lying states in $^{196-204}$Hg. Quantitatively some theoretical results fit somewhat worse. Especially for the value from the $2_{2}^{+}$ state to the $2_{1}^{+}$ state, the experimental one is much smaller. These deficiencies are understandable. $^{198-204}$Hg are near the magic nuclei and their proton number is 80. Compared with the B(E2) value from the $2_{2}^{+}$ state to the $2_{1}^{+}$ state in $^{198}$Pt, that is 37(7) W.u., the very small value in $^{198}$Hg looks strange. The theoretical results of the $2_{2}^{+}$ states in Fig. 2 also suffer from deficiencies. These may indicate that the collective nature of the $2_{2}^{+}$ state is greatly reduced. This reason is worth studying. First, it comes from the mixing of the single-particle excitation, which can be studied by the future proxy-SU(3) shell model \cite{Bonatsos171,Bonatsos23} with SU(3) higher-order interactions. Second, in the proxy-SU(3) symmetry, the last two single particle orbits (proton number 81 and 82) can not be included \cite{Bonatsos20}, thus the SU(3) symmetry may be partly broken. Thus $^{196-204}$Hg (also $^{192-200}$Pt) provide an ideal platform to discuss various possible mechanisms, such as the SU3-IBM, adding the $g$ boson, distinguishing the protons and neutrons, the proxy-SU(3) symmetry, the mapping from the shell model SU(3) to the boson SU(3) \cite{Bonatsos86, Elliott99}.

The experimental and theoretical quadrupole moments $Q_{2_{1}^{+}}$ of the $2_{1}^{+}$ states in $^{196-204}$Hg are shown in Fig. 4. The theoretical values fit well for $^{198,204}$Hg. For $^{200}$Hg, the experimental value is nearly twice of the theoretical one. For $^{202}$Hg, one value is nearly twice while another (red triangle) is nearly the same as the theoretical value. We notice that the theoretical results are similar to the ones in $^{196,198}$Pt. The fitting is reasonable.

$^{196-204}$Hg have been studied in Ref. \cite{Bernards131,1,2,3,Nomura13,Heyde14,Utsuno24,6,Chakraborty19,8,9,Obaizola19}, but the boson number odd-even effect are not observed. In \cite{Bernards132}, this effect was noticed, and they thought that it is an anomalous phenomenon. In \cite{Heyde14,Utsuno24}, $^{196-200}$ Hg were explained at a good level based on the O(6)-U(5) evolution, however this explanation can not be generalized to $^{202,204}$Hg because the $0_{2}^{+}$ state in $^{202}$Hg moves up again which is not a U(5) nucleus. As far as we know, no theory has involved this odd-even effect and explained it. As our theory suggests, this interpretation involves the SU(3) symmetry and the higher-order interactions, which goes beyond the traditional idea of understanding the shape and evolution of nuclear structures. Thus previous theories are also unlikely to provide an explanation.

From the discussions above, the idea that the SU(3) third-order Casimir operator describes the oblate shape is proved correct. Thus the description with the $\overline{\textrm{SU(3)}}$ symmetry is not in line with the actual situation. This conclusion has been preliminarily seen in pevious work \cite{wang23}. In this Letter, we present evidence for certainty. The O(6) symmetry is the critical point from the prolate shape to the oblate shape described by the $\overline{\textrm{SU(3)}}$ symmetry \cite{Wang08,Werner01,Jolie02,Warner02,Jolie031,Bonatsos24}, so this symmetry is also unreasonable to describe the $\gamma$-softness of the realistic nuclei. The O(6) symmetry has been found not to describe the B(E2) anomaly \cite{wangtao}.

In conclusion, the SU3-IBM predicts the boson number odd-even effect for nuclei, which is found to really exist in $^{196-204}$Hg. We explain this effect very well with a single Hamiltonian in the SU3-IBM, which provides a conclusive evidence for the accuracy and validity of the SU3-IBM. Because these results are very sensitive to the boson number $N$, they also verify the basic boson number hypothesis of the IBM, which was considered impossible in the past 50 years. In this study, we show that, the SU(3) symmetry and the higher-order interactions are both important for understanding the complicated low-lying behaviors in nuclei. SU(3) third-order Casimir operator and other hgiher-order interactions are essential for understanding the oblate shape. Further experimental explorations on the Hg isotopes and various extensions of the SU3-IBM are needed to further verifying the deterministic relationship found in this Letter.

\end{document}